\documentstyle[12pt]{article}

\begin{document}
\begin{center}
{\Large \bf
Quantum mechanical constraints on the measurement
of the density of the electromagnetic energy}

\bigskip

{\large D.L.~Khokhlov}
\smallskip

{\it Sumy State University, R.-Korsakov St. 2, \\
Sumy 40007, Ukraine\\
E-mail: khokhlov@cafe.sumy.ua}
\end{center}

\begin{abstract}

It is considered constraints imposed by the quantum mechanics
on the measurement of the density of the electromagnetic
energy. First, the energy of the electromagnetic wave and the
volume (time) are bound with the Heisenberg uncertainty
principle. It reduces from double to ordinary
the relativistic effect for the density of the electromagnetic
energy. Second, the frequency of photons and the number of photons
are bound with the Heisenberg uncertainty principle.
Then relativistic effects for the frequency of photons and for the
number of photons are incompatible.

\end{abstract}

The energy density of the electromagnetic wave is given by
\begin{equation}
W=\frac{E}{V}=\frac{E}{St}
\label{eq:W}
\end{equation}
where $E$ is the energy of the electromagnetic wave, $V$ is the
volume, $S$ is the cross-section, $t$ is the time.
Let the source of the electromagnetic wave move with the velocity
$v$ relative to the laboratory frame. Let an observer in the
laboratory frame measure the energy density of the electromagnetic
wave. According to the special
relativity~\cite{Lan}, the energy of the
electromagnetic wave and the volume (the time) follow the Lorentz
transformation. Then the energy of the
electromagnetic wave and the volume (the time) measured in the
laboratory frame are shifted with respect to those in the frame of
the source. When measuring the energy density of the
electromagnetic wave emitted by the moving source, we obtain the
double relativistic effect
\begin{equation}
W^\prime=\frac{W(1-v^2/c^2)}{(1\pm v/c)^2}.
\label{eq:W2}
\end{equation}

Electromagnetic wave is a quantum object. Then we should take into
account quantum mechanical constraints on the measurement of the
density of the electromagnetic energy.
Due to the Heisenberg uncertainty principle we cannot
simultaneously
measure the momentum and the space coordinate (the energy
and the time) of the electromagnetic wave. When determining the
energy of the electromagnetic wave emitted in the moving frame,
we cannot determine the volume (the time) in the moving frame and
should use the volume (the time) determined in the laboratory
frame. Then, the energy of the
electromagnetic wave is shifted in accordance with the Lorentz
transformation, but the volume (the time) is taken as a
laboratory one. As a result we obtain the
ordinary relativistic effect for the density of the
electromagnetic energy
\begin{equation}
W^\prime=\frac{W(1-v^2/c^2)^{1/2}}{1\pm v/c}.
\label{eq:W3}
\end{equation}
Thus, due to quantum mechanical constraints, the conventional
formula of the relativistic effect for the density of the
electromagnetic energy given by eq.~(\ref{eq:W2}) is not correct.
Instead of this we should use the formula given by
eq.~(\ref{eq:W3}).

Consider electromagnetic wave as a bunch of photons with the
frequency $\omega$. The luminosity of the bunch of photons is
given by
\begin{equation}
L=\frac{E}{St}=\frac{n\hbar\omega}{St}
\label{eq:L}
\end{equation}
where $n$ is the number of photons, $\hbar$ is the Planck
constant. The luminosity of the bunch of photons emitted by the
moving source suffers the ordinary relativistic shift given by
eq.~(\ref{eq:W3}).
In the bunch of photons, the frequency and the number of photons
are bound with the Heisenberg uncertainty principle. Hence we
cannot simultaneously determine the frequency of photons
and the number of photons. Hence relativistic effects for the
frequency of photons and for the number of photons can be measured
in the different incompatible experiments.
When measuring the frequency of photons emitted by the moving
source, the number of photons is taken as a laboratory one, and
the frequency of photons suffers the Doppler shift,
$\omega\propto(1-v^2/c^2)^{1/2}/(1\pm v/c)$.
When measuring the luminosity of the bunch of photons,
the frequency of photons is taken as a laboratory one,
and the number of photons is shifted as
$n\propto(1-v^2/c^2)^{1/2}/(1\pm v/c)$.
Under the conventional approach, when measuring the luminosity
of the bunch of photons in the moving frame, we determine both the
Doppler shift of the photon's frequency and the shift of the
number of photons. However this contradicts to the Heisenberg
uncertainty principle.

Thus the quantum mechanics, the Heisenberg uncertainty principle,
imposes constraints on the measurement of the density of the
electromagnetic energy. First, it reduces from double to ordinary
the relativistic effect for the density of the electromagnetic
energy. Second, relativistic effects for the frequency of photons
and for the number of photons are incompatible.


\begin{thebibliography}{99}

\bibitem{Lan}
L.D. Landau and E.M. Lifshitz, {\it The classical theory of
fields}, 4th Ed. (Pergamon, Oxford, 1976).

\end{thebibliography}
\end{document}